\documentclass[runningheads]{llncs}
\usepackage[T1]{fontenc}
\usepackage{graphicx}
\begin{document}
\title{Fast-MC-PET: A Novel Deep Learning-aided Motion Correction and Reconstruction Framework for Accelerated PET}
\titlerunning{Deep Learning-aided Accelerated PET Reconstruction}
%
\author{Bo Zhou\inst{1} \and Yu-Jung Tsai\inst{2} \and Jiazhen Zhang\inst{1} \and Xueqi Guo\inst{1} \and Huidong Xie\inst{1} \and Xiongchao Chen\inst{1} \and Tianshun Miao\inst{2} \and Yihuan Lu\inst{3} \and James S. Duncan\inst{1,2} \and \\ Chi Liu\inst{1,2}}
\authorrunning{B. Zhou, et al.}
%
%

\institute{Department of Biomedical Engineering, Yale University
\and
Department of Radiology and Biomedical Imaging, Yale University \and United Imaging Healthcare \\
\email{bo.zhou@yale.edu}
}

\maketitle              
\begin{abstract}
Patient motion during PET is inevitable. Its long acquisition time not only increases the motion and the associated artifacts but also the patient's discomfort, thus PET acceleration is desirable. However, accelerating PET acquisition will result in reconstructed images with low SNR, and the image quality will still be degraded by motion-induced artifacts. Most of the previous PET motion correction methods are motion type specific that require motion modeling, thus may fail when multiple types of motion present together. Also, those methods are customized for standard long acquisition and could not be directly applied to accelerated PET. To this end, modeling-free universal motion correction reconstruction for accelerated PET is still highly under-explored. In this work, we propose a novel deep learning-aided motion correction and reconstruction framework for accelerated PET, called Fast-MC-PET. Our framework consists of a universal motion correction (UMC) and a short-to-long acquisition reconstruction (SL-Reon) module. The UMC enables modeling-free motion correction by estimating quasi-continuous motion from ultra-short frame reconstructions and using this information for motion-compensated reconstruction. Then, the SL-Recon converts the accelerated UMC image with low counts to a high-quality image with high counts for our final reconstruction output. Our experimental results on human studies show that our Fast-MC-PET can enable 7-fold acceleration and use only 2 minutes acquisition to generate high-quality reconstruction images that outperform/match previous motion correction reconstruction methods using standard 15 minutes long acquisition data. 

\keywords{Accelerated PET \and Universal Motion Correction \and Deep Reconstruction.}
\end{abstract}

\section{Introduction}
\vspace{-0.18cm}
Positron Emission Tomography (PET) is a commonly used functional imaging modality with wide applications in oncology, cardiology, neurology, and biomedical research. However, patient motion during the PET scan, including both involuntary motions (i.e. respiratory, cardiac, and bowel motions) and voluntary motions (i.e. body and head motions), can lead to significant motion artifacts, degrading the downstream clinical tasks. Moreover, the long acquisition time that easily exceeds 15 minutes, will lead to increased patient motion, patient discomfort, and low patient throughput. 

In previous works of PET motion correction (MC), a variety of external device-aided and data-driven MC methods have been developed for correcting specific motion types. For example, in respiratory MC, Chan \emph{et al.} \cite{chan2017non} developed a non-rigid event-by-event continuous MC list-mode reconstruction method. Lu \emph{et al.} \cite{lu2018respiratory} further improved their method by generating matched attenuation-corrected gate PET for respiratory motion estimation. In body MC, Andersson \emph{et al.} \cite{andersson1998obtain} proposed to divide the PET list-mode data into predefined temporal frames for reconstructions, where the reconstructions of each frame are registered to a reference frame for body MC. Later, Lu \emph{et al.} \cite{lu2019data} further developed a reconstruction-free center-of-distribution-based body motion detection and correction method. In cardiac MC, cardiac cycle tracking/gating using electrocardiography (ECG) is still the gold-standard \cite{lu2020patient}. While providing efficient MC solutions to reduce motion artifacts for different motion types, these methods usually require prior knowledge of the motion type and need motion-type-specific modeling. Thus, these previous MC methods may lead to sub-optimal image quality or fail when multiple motion types are present simultaneously. There are also recent attempts in using ultra-fast list-mode reconstruction of short PET frames to estimate motion during the PET scan \cite{spangler2021ultra,zhang2021deep}. However, these methods may not adapt well to many motion types with non-rigid motion \cite{spangler2021ultra}, and extending to non-rigid motion is computationally infeasible, i.e. requiring non-rigid registration of thousands of frames for a single scan using traditional registration algorithms \cite{zhang2021deep}. In addition, it still requires the standard long acquisition to collect sufficient events to achieve a reasonable signal-to-noise ratio (SNR) in the final reconstruction. On the other hand, previous works have also investigated the feasibility of reducing the PET acquisition time. Lindemann \emph{et al.} \cite{lindemann2018towards} and Lasnon \emph{et al.} \cite{lasnon2020fast} found that one can reasonably maintain the PET image quality and lesion detectability with two-fold acquisition time reduction using traditional reconstructions. Weyts \emph{et al.} \cite{weyts2022artificial} show that a deep learning-based denoising model can enable two-fold PET acquisition time reduction and provide image quality that matches with the full acquisition. However, these works only show the feasibility of a 2-fold time reduction and did not consider the residual motions during the accelerated acquisition. 

In this work, we aim to address these challenges by developing a PET reconstruction framework that can 1) reduce the acquisition time, i.e. 7-fold acceleration, and 2) correct the residual motion, regardless of the motion type, in the accelerated acquisition. Specifically, we propose a novel deep learning-aided data-driven motion reduction and accelerated PET reconstruction framework, called Fast-MC-PET. In the Fast-MC-PET, we first design a universal motion correction method aided by deep learning to reconstruct a motion-reduced image from the short acquisition. While reducing the motion artifacts given the accelerated acquisition and our motion correction, the reconstructed image still suffers from high noise levels due to low event counts. Thus, in the second step of Fast-MC-PET, we also deploy a deep generative network to convert the low-counts images to high-counts images. Our experimental results on real human data demonstrate the Fast-MC-PET can generate high-quality images with reduced motion-induced errors while enabling 7-fold accelerated PET acquisition. 

\vspace{-0.18cm}
\section{Methods}
\vspace{-0.18cm}
Our Fast-MC-PET consists of two key components, including a universal motion correction (UMC) module and a short-to-long acquisition reconstruction (SL-Recon) module. In UMC, we first partition the list-mode data into ultra-short list-mode data, i.e. every 500ms, and estimate a quasi-continuous motion over the short acquisition. Given the motion and the original list-mode data, a motion-corrected short-acquisition image is then reconstructed by a motion-compensated OSEM list-mode reconstruction. Finally, a deep generative model is devised to transform the motion-corrected short-acquisition image into a high-count long-acquisition image, thus providing a motion-corrected high-count image using only accelerated short-acquisition. In the following sections, we will describe these steps in detail.

\vspace{-0.45cm}
\begin{figure*}[htb!]
\centering
\includegraphics[width=0.96\textwidth]{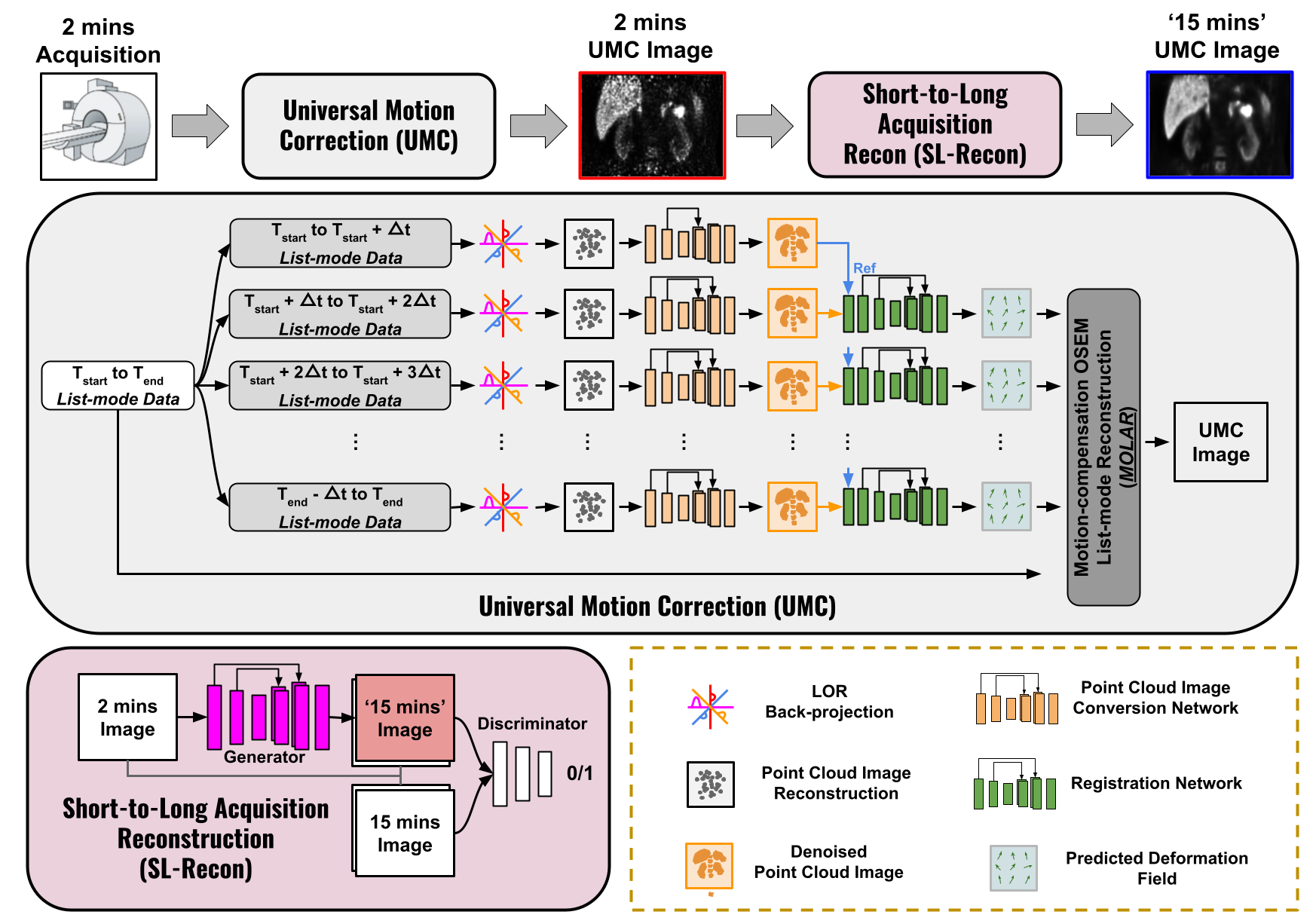}
\caption{The overall pipeline of Fast-MC-PET. The Universal Motion Correction (UMC) module (grey box) reconstructs motion-reduced image from the short acquisition data. The Short-to-Long Acquisition Reconstruction (SL-Recon) module (pink box) converts the UMC image from short acquisition to long acquisition.}
\label{fig:pipeline}
\end{figure*}

\subsection{Universal Motion Correction}
\vspace{-0.15cm}
With the short acquisition data, the UMC aims to generate a motion reduced low-count reconstruction. The UMC consists of three steps, including point cloud image (PCI) \& paired gated image generation, quasi-continuous motion estimation, and motion-compensated OSEM list-mode reconstruction. \\

\vspace{-0.20cm}
\noindent\textbf{Point Cloud \& Paired Gated Image Generation.} To estimate a continuous motion, the list-mode data is first partitioned into a series of ultra-short list-mode data, i.e. every 500ms. For every 500ms list-mode data, we back-project the Line-of-Response (LOR) of each event within the time-of-flight (TOF) bin, and all the back-projected LORs form a PCI for this short time frame. The PCI reconstruction can be formulated as
\begin{equation}
    P_{j,t} = \sum_i \frac{ c_{i,j,t} L_{i,t} }{ Q_j },
\end{equation}
where $c_{i,j,t}$ is the system matrix that represents the contribution of an
annihilation originating from pixel $j$ being detected on LOR $i$ at time $t$, accounting for geometry, resolution, and solid angle effects. $L_{i,t}$ is the decay correction factor. $Q_j$ is the sensitivity of voxel $j$ that is pre-computed via $Q_j = \sum_i c_{i, j}$, and $P_{j,t}$ is the back-projected value of voxel $j$ at time $t$ with sensitivity correction. 

\begin{figure*}[htb!]
\centering
\includegraphics[width=0.99\textwidth]{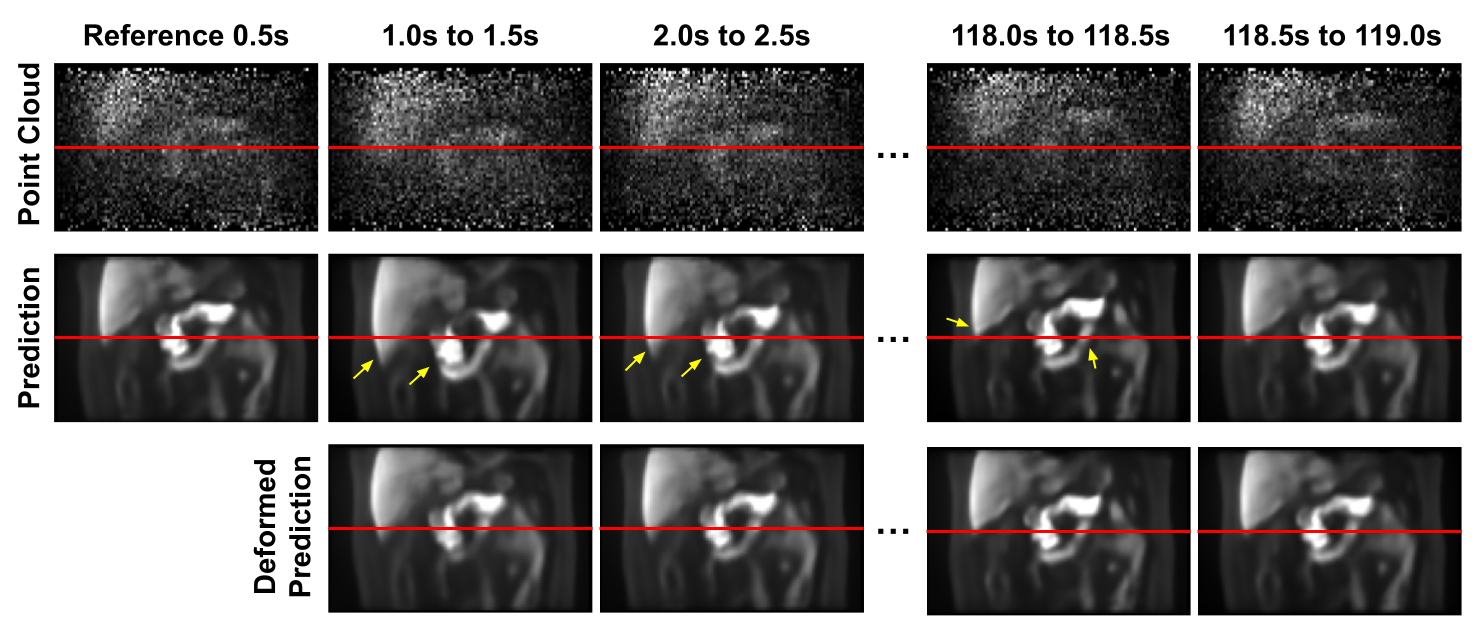}
\caption{Examples of the Point Cloud Images (PCIs), the denoised PCIs (dPCIs), and the deformed dPCIs using estimated motion fields.}
\label{fig:pc}
\end{figure*}

Due to the ultra-low-counts level, the signal-to-noise ratio (SNR) of PCI is low and is unsuitable for motion estimation tasks, as demonstrated in Figure \ref{fig:pc}'s 1st row. Thus, we deploy a deep learning-based denoising network, i.e. UNet \cite{ronneberger2015u}, that aims to convert PCIs to gated OSEM images with high SNR. To train the denoising network, we first reconstruct the amplitude-based respiratory gated OSEM images \cite{ren2017data} using the body motion free list-mode data, extracted by the Centroid-Of-Distribution (COD)-based body motion detection method \cite{lu2019data}. Then, within each gate, we randomly extract 10\% PCIs to construct the training pairs of PCI and the corresponding gated image. $\mathcal{L}_2$ loss is used for the network training, and can be formulated as
\begin{equation}
\mathcal{L}_{dn} = || \gamma_g - f_{dn}(P_g) ||_2^2
\end{equation}
where $\gamma_g$ is the gated OSEM image and $P_g$ is the randomly extracted PCI that lies in the same gate. With a trained denoising model $f_{dn} (\cdot)$, the series of PCIs can then be converted to a series of high-quality denoised PCI (dPCI) via:
\begin{equation}
    \gamma_t = f_{dn}(P_t)
\end{equation}
where $\gamma_t$ is the denoised images with $t = (0\sim\Delta t, \Delta t \sim 2\Delta t, ..., T-\Delta t \sim T)$. Here, we set $\Delta t = 0.5s$ and $T = 120s$ here, thus generating 240 3D images. Examples of dPCIs are illustrated in Figure \ref{fig:pc}'s 2nd row. \\

\vspace{-0.20cm}
\noindent\textbf{Quasi-continuous Motion Estimation.} A quasi-continuous motion can be estimated using the series of dPCIs from the previous step. Within the first 5 seconds, the dPCI in the expiration phase, i.e. with the highest COD coordinates in the z-direction, is chosen as the reference frame $\gamma_{ref}$ for all the other frames $\gamma_t$, resulted in 239 dPCI pairs requiring registration. Conventional registration methods \cite{xu2016evaluation,papademetris2006bioimage} are time-consuming, and it is prohibitively long to register hundreds of 3D pairs here. Thus, we propose to use a deep learning-based registration method for fast motion estimation \cite{balakrishnan2019voxelmorph} in our framework. Given the reference dPCI image $\gamma_{ref}$ and the source dPCI image $\gamma_t$, we use a motion estimation network, i.e. UNet \cite{ronneberger2015u}, to predict the motion deformation $M_t = f_{m} (\gamma_{ref}, \gamma_t)$. The network is trained by optimizing the following loss function: 
\begin{equation}
    \mathcal{L}_m = ||\gamma_{ref} - M_t \circ \gamma_t||_2^2 + \beta || \nabla M_t ||_2^2
\end{equation}
where the first term measures the image similarity after applying the motion prediction $M_t$, and the second term is a deformation regularization that adopts a L2-norm of the gradient of the deformation. The regularization's weight is set as $\beta=0.001$. During training, $\gamma_{ref}$ and $\gamma_t$ are randomly selected from the gated images. With a trained motion estimation network $f_m(\cdot)$, we can then estimate the quasi-continuous motion using $M_t = f_{m} (\gamma_{ref}, \gamma_t)$ with $t = (0\sim\Delta t, \Delta t \sim 2\Delta t, ..., T-\Delta t \sim T)$. \\

\vspace{-0.20cm}
\noindent\textbf{Motion-compensated OSEM List-mode Reconstruction.} To reconstruct a single image $\lambda$ at the reference location $\gamma_{ref}$ using all the coincidence events, we can deform the system matrix at each time $t$ to the reference location, generating new deformed system matrixs $c_{i,j}^{t \rightarrow ref}$ using $M_t$ from the previous step. Deforming the system matrix can be seen as "bending" the LORs into curves of response (CORs), where both forward and back-projections are traced along the CORs. In list-mode notation, for event $k$ occurring on LOR $i(k)$ at time $t(k)$, we replace indexes $i$ by $k$, and substitute $c_{k,j}$ in the previous TOF-MOLAR \cite{jin2013list} by $c_{k,j,\tau_k}^{t \rightarrow ref}$. The OSEM updating equation can thus be formulated as:
\begin{equation}
    \lambda_j^{n+1} = \frac{ \lambda_j^{n} }{ Q_j } \sum_{k=1}^K \frac{c_{k,j,\tau_k}^{t \rightarrow ref} L_k A_k N_k}{T( \sum_{j'} c_{k,j',\tau_k}^{t \rightarrow ref} L_k A_k N_k \lambda_{j'}^{n} + R_{k,\tau_k} + S_{k,\tau_k})}
\end{equation}
\begin{equation}
    Q_j = \frac{1}{n_T} \sum_{t'=1}^{n_T} \sum_{i=1}^{I} \sum_{\tau=1}^{n_\tau} c_{i,j,t',\tau}^{t \rightarrow ref} L_{i,t'} A_{i,t'} N_i
\end{equation}
where $n$ is the number of iteration, $k$ is the index of each detected event, $c_{k,j,\tau_k}^{t \rightarrow ref}$ is the deformed system matrix element with $\tau_k$ denoting the TOF bin for event $k$. $L_k$ is the decay factor and $A_k$ is the attenuation factor derived from CT. $N_k$ is the sensitivity term, $R_{k,\tau_k}$ is the randoms rate estimate, and $S_{k,\tau_k}$ is the scatter rate estimate in counts per second in TOF bin $\tau_k$. The random events are estimated from the product of the singles rates of the two detectors for each LOR, and then uniformly distributed across all TOF bins. Here, $Q$ is the sensitivity image that is pre-computed by back-projecting randomly sampled events along the CORs to account for the motion on voxel sensitivity. When calculating $Q$, each time frame of duration $T$ is divided into $n_T$ short time bins, i.e. $t'$. Moreover, $n_\tau$ denotes the total number of TOF bins ($n_\tau=13$ for the Siemens mCT PET scanner used in this study). Here, we set the number of iteration to 2 and the number of subsets to 21 for our UMC reconstructions. 

\vspace{-0.25cm}
\subsection{Short-to-Long Acquisition Reconstruction}
\vspace{-0.15cm}
Even though the UMC reduces the motion effects in the reconstruction, the UMC image still suffers from low SNR due to the limited counts from the short acquisition, as compared to the long acquisition. Thus, we propose to use a short-to-long acquisition reconstruction (SL-Recon) to convert the UMC image from a short-acquisition to a long-acquisition one. Here, we use a conditional generative adversarial network for this reconstruction. Given a UMC image $\lambda_s$ from the short acquisition, we can use a generative network, i.e. UNet \cite{ronneberger2015u}, that directly predicts the UMC image $\lambda_l$ from a long acquisition from it. The SL-Recon network is trained using both a pixel-wise L2 loss and an adversarial loss defined as:
\begin{equation} 
    \mathcal{L}_{2} = || G(\lambda_s) - \lambda_{l} ||_2^2
\end{equation}
\begin{equation}
    \mathcal{L}_{adv} = - log (D_{gan}(\lambda_l | \lambda_s)) - log (1 - D_{gan}(G(\lambda_s) | \lambda_s))
\end{equation}
where $G$ is the SL-Recon generative network and $D$ is the discriminator network. Here, we simply use OSEM reconstructions from long acquisitions (15 minutes), paired with OSEM reconstructions from short acquisitions (2 minutes in the center period), for the network's training.  

\vspace{-0.25cm}
\subsection{Evaluation on Human Data}
\vspace{-0.15cm}
We included 26 pancreatic \textsuperscript{18}F-FPDTBZ \cite{normandin2012vivo} PET/CT patient studies. All PET data were obtained in list mode using the 4-ring Siemens Biograph mCT scanners equipped with the AZ-733V respiratory gating system (Anzai Medical, Tokyo, Japan). The Anzai respiratory trace was recorded at 40 Hz for all subjects. The average dose administered to the patients is 9.13±1.37 mCi. $15$ minutes of the list-mode acquisition were used for each patient study. We used 23 patients to generate the training data for the PCI denoising model, the motion estimation model, and the SL-Recon model. Extensive evaluations were performed on the remaining 3 patients with different motion types. For training the PCI denoising model and the motion estimation model, we generated 5 gated images for each patient using OSEM (21 subsets and 2 iterations). For training the SL-Recon model, the training pairs of long/short acquisition images were reconstructed using the same OSEM protocol without gating. All the images were reconstructed into $200 \times 200 \times 109$ 3D volumes with a voxel size of $2.032 \times 2.032 \times 2.027$ $mm^3$. 

\vspace{-0.25cm}
\subsection{Implementation Details}
\vspace{-0.15cm}
We implemented our deep learning modules using Pytorch. We used the ADAM optimizer \cite{kingma2014adam} with a learning rate of $10^{-4}$ for training the PCI denoising network, motion estimation network, and the SL-Recon network. We set the batch size to 3 for all networks' training. All of our models were trained on an NVIDIA Quadro RTX 8000 GPU. The PCI denoising network was trained for 200 epochs, and then fine-tuned for 10 epochs on the patient-specific gated images of the test patient during the test time. The motion estimation network was trained for 250 epochs, and the SL-Recon network was trained for 200 epochs. To prevent overfitting, we also implemented 'on-the-fly' data augmentation for the PCI denoising and SL-Recon networks. During training, we performed $64 \times 64 \times 64$ random cropping, and then randomly flip the cropped volumes along the x, y, and z-axis.

\begin{figure*}[htb!]
\centering
\includegraphics[width=0.99\textwidth]{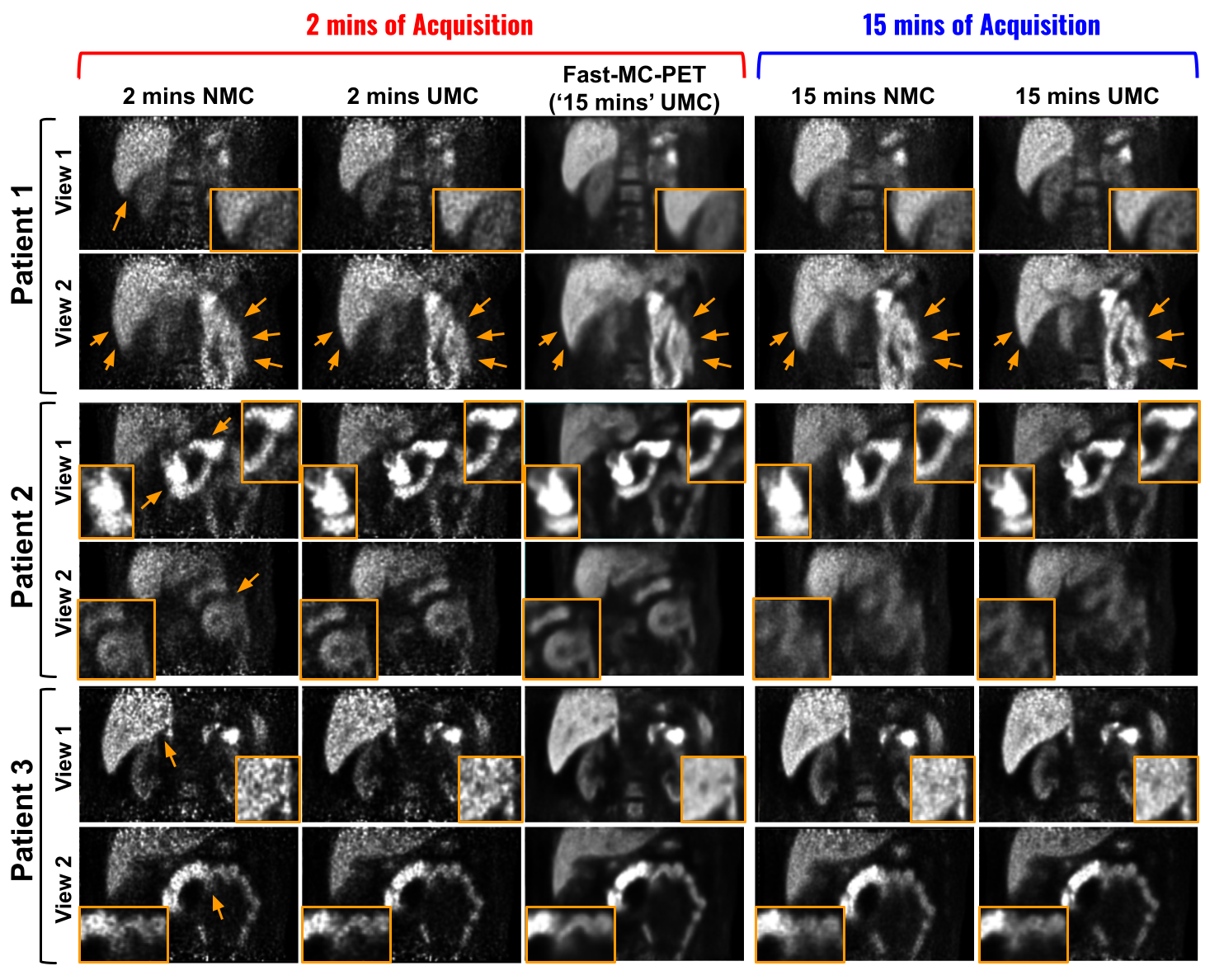}
\caption{Visualization of Fast-MC-PET reconstructions. The 2min UMC images (2nd column) contain less motion blurring, as compared to the no motion correction (NMC) images (1st column). The virtual 15 minutes UMC images (3rd column) predicted from 2 minutes UMC images (2nd column) provide image-quality that match the true 15 minutes images (last column).}
\label{fig:2to15}
\end{figure*}

\vspace{-0.15cm}
\section{Results}
\vspace{-0.25cm}
The qualitative comparison of Fast-MC-PET reconstructions is shown in Figure \ref{fig:2to15}. As we can observe, the 2 minutes reconstruction with no motion correction (NMC) suffers from both motion blurring and high-noise levels due to low counts. The first patient has both body/torso motion and respiratory motion during the 2 minutes PET scan, thus introducing heavy blurring for major organ boundaries, i.e. liver and kidneys. The 2 minutes UMC image recovers the sharp organ boundaries by correcting those motions during the short acquisition. Based on the UMC image from 2 minutes acquisition, the final Fast-MC-PET image further reduces the noise thus providing a near motion-free and high-count image, matching the 15 minutes UMC image quality. The second patient with respiratory and bowel motion introduces significant image blurring for the pancreas (view 1) and intestines (view 2). The 2 min UMC image can recover the diminished details inside these organ regions. The final Fast-MC-PET image further reduces the noise, thus generating a high-quality image with motion correction and high counts. On the other hand, by reducing the acquisition time from 15 minutes to 2 minutes, we can see that the diminished organ structures, especially the intestine structure (view 2) in 15 minutes NMC, can be preliminarily restored in 2 minutes NMC. Complex motion, e.g. bowel motion, in a 15 minutes long acquisition is extremely challenging to correct even with UMC. Thus, based on 2 minutes acquisition, the Fast-MC-PET here shows better reconstruction quality with better structural recovery. Similar observations can be found for the third patient with respiratory and bowel motion, where the 2 minutes-based Fast-MC-PET provides reconstruction quality matched the 15 minutes UMC reconstruction. 

\begin{figure*}[htb!]
\centering
\includegraphics[width=0.99\textwidth]{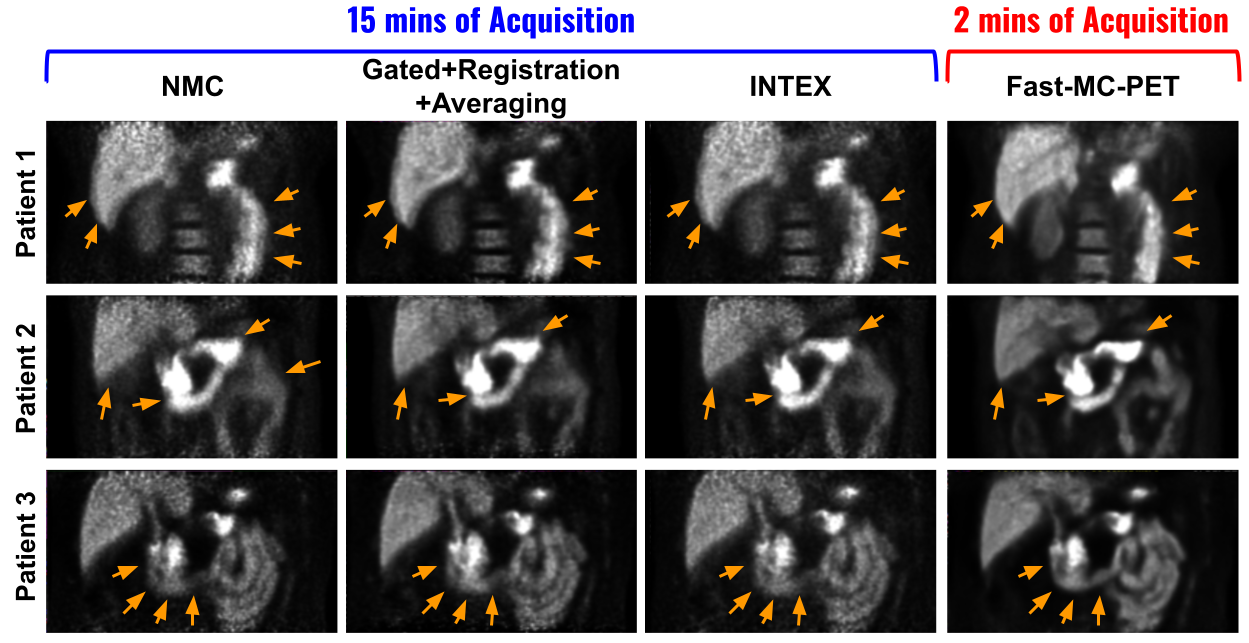}
\caption{Comparisons to previous motion correction methods. Our Fast-MC-PET with 2 minutes acquisition show improved structural details recovery (orange arrows), as compared to previous methods with 15 minutes acquisition.}
\label{fig:comp}
\end{figure*}

We compared our 2 minutes-based Fast-MC-PET reconstructions to previous correction methods that are long acquisition based, i.e. 15 minutes. The visual comparison is shown in Figure \ref{fig:comp}. First, we compared with the classic respiratory motion correction method \cite{bai2009regularized} that reduces the motion and noise by averaging the aligned amplitude-gated images, where non-rigid registration \cite{papademetris2006bioimage} is used for alignments. Then, we compared our method with the NR-INTEX \cite{chan2017non} that compensates for the respiratory motion by estimating the continuous deformation field using internal-external motion correlation which is considered the current state-of-the-art method. Both previous methods require specific motion-type modeling, and thus fail when additional motion types are present, e.g. body motion (Patient 1) and bowel motion (Patient 3). The UMC module in the Fast-MC-PET is not specific to any motion type and thus can correct different types of motion together. Therefore, our Fast-MC-PET can provide consistently better results when multiple types of motion co-exist (Patients 1 and 3), and generate comparable reconstruction quality when respiratory motion is dominating (Patient 2). 

\begin{figure*}[htb!]
\centering
\includegraphics[width=0.99\textwidth]{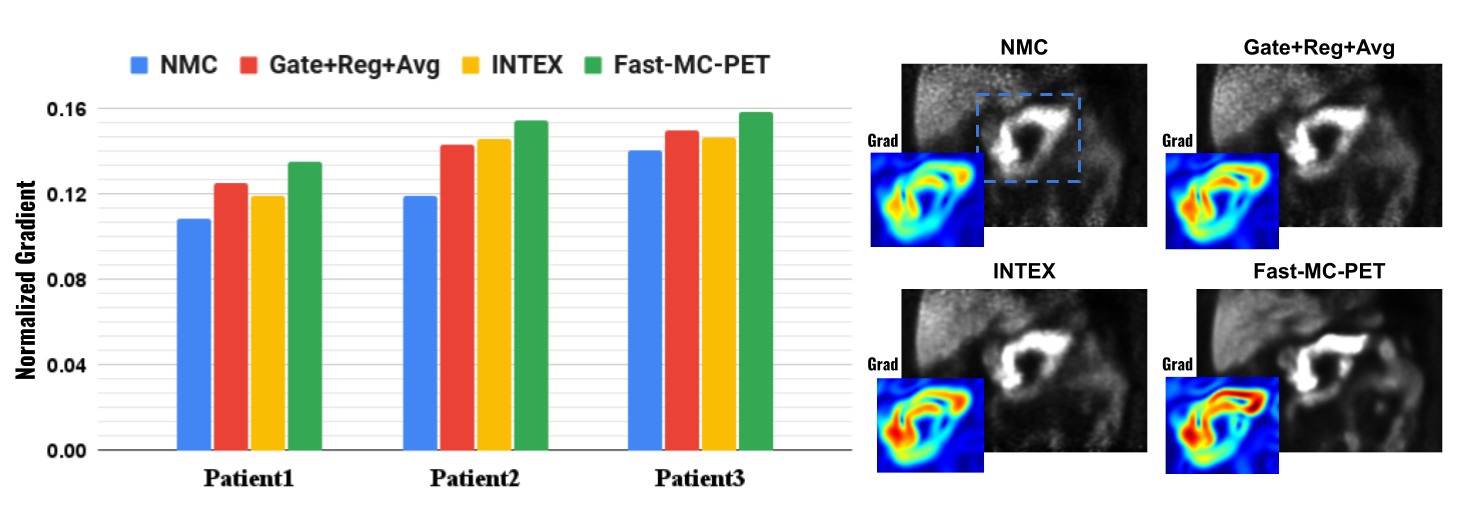}
\caption{Comparison of the gradient of reconstructions. Left: quantitative evaluation using the mean gradient value. Right: visual comparison of the reconstruction and the gradient.}
\label{fig:grad}
\end{figure*}

For quantitative evaluation, we computed the mean normalized gradient of the reconstructions, where better reconstruction with sharper structure will have higher gradient values. The results are summarized in Figure \ref{fig:grad}. The normalized gradient values of Fast-MC-PET are $0.159$, $0.154$, and $0.132$ for Patients 1, 2, and 3, respectively, which are consistently higher than all previous methods. A comparison example from Patient 2 is shown on the right. The gradient image of the Fast-MC-PET using only 2 minutes acquisition shows higher gradient values and more continuous structure patterns when compared to previous methods based on 15 minutes acquisition. 

\begin{figure*}[htb!]
\centering
\includegraphics[width=0.99\textwidth]{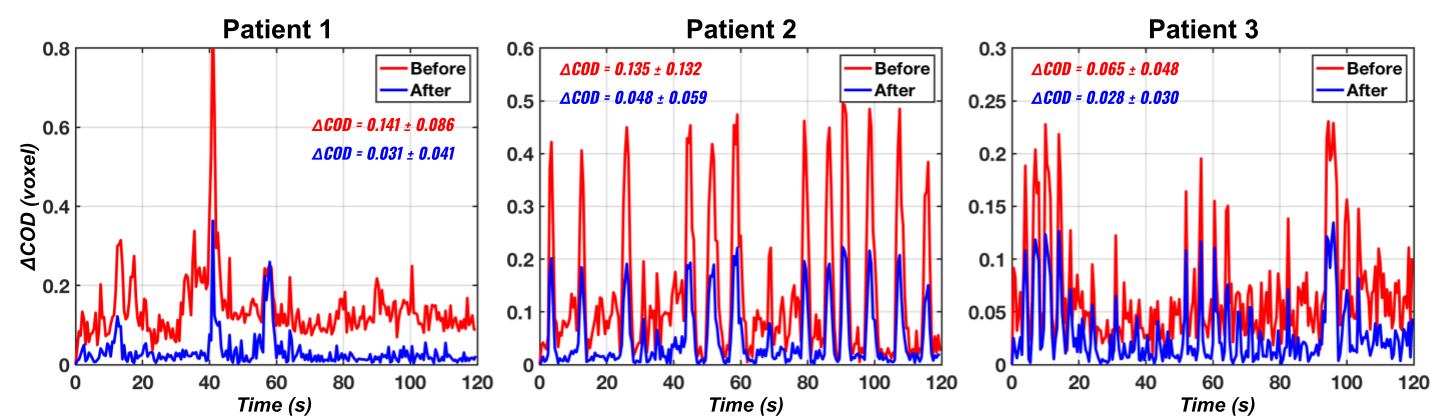}
\caption{The difference of COD trace between the reference frame and the current frame ($\Delta$COD) over the 2 minutes acquisition. The $\Delta$COD before (red) and after (blue) UMC correction are plotted for all three patients. The mean $\Delta$CODs are reported in the plots.}
\label{fig:cod}
\end{figure*}

Ablative evaluation of motion correction is shown in Figure \ref{fig:cod}. The difference of COD between the reference frame and the current frame ($\Delta$COD) over the 2 minutes acquisition is visualized. For Patient 1 with body motion and irregular breathing pattern, the $\Delta$COD curve before correction contains irregular steep changes leading to a mean $\Delta$COD of $0.141 \pm 0.086$. With the UMC in our Fast-MC-PET, the curve after correction is much more stable with a reduced mean $\Delta$COD of $0.031 \pm 0.041$ with significance ($p < 0.001$). For Patients 2 and 3 with more stable and regular motion patterns, the UMC can also reduce the mean $\Delta$COD from $0.135 \pm 0.132$ to $0.048 \pm 0.059$ and from $0.065 \pm 0.048$ to $0.028 \pm 0.030$, respectively. Both with significance ($p < 0.001$). A patient example of PCIs over the 2 minutes acquisition before and after applying the UMC correction is shown in Figure \ref{fig:pc}.

\vspace{-0.15cm}
\section{Discussion}
\vspace{-0.25cm}
In this work, we propose a novel deep learning-aided data-driven motion correction and reconstruction framework for accelerated PET (Fast-MC-PET). The proposed method can accelerate the PET acquisition by nearly 7-fold and use only 2 min acquisition while providing high-quality reconstruction with motion correction. In this framework, we first devise a UMC module that estimates continuous motion based on PCIs and use this information to reconstruct motion-compensated images. Instead of using 15 minutes long acquisition that 1) inherits more motion due to long scanning time and 2) requires registrations of $1800$ PCI pairs in UMC, we use 2 minutes accelerated acquisition with less motion and only requires registrations of $240$ PCI pairs. The averaged registration inference time for one pair is $0.41$s, thus needing about $98.5$s for all registration in UMC which is more manageable. The UMC reconstruction from accelerated acquisition can then be inputted into the SL-Recon module to directly generate the 15 minutes long acquisition motion corrected reconstruction. With this simple yet efficient pipeline, we can generate high-quality motion corrected accelerated PET reconstruction that potentially outperforms previous methods with the standard long acquisition. 

There are a few limitations and opportunities that are the subject of our ongoing work. First, our pilot study only test on \textsuperscript{18}F-FPDTBZ patients who were all scanned using Siemens mCT. The trained model may not directly generalize well to a different PET tracer/scanner. However, if the training data of different tracers/scanners is available, the Fast-MC-PET can be fine-tuned and potentially adapted to these distributions. Multi-institutional federated learning \cite{zhou2022federated} may also be used to improve the adaptation. In the future, we will further evaluate the performance using patients scanned with different PET tracers/scanners. Second, we used a temporal resolution of 500 ms for PCI in UMC with a focus on abdominal region motion correction in this work. A higher temporal resolution, e.g. 100ms, may be needed for cardiac motion correction in the chest region, which is an important direction in our future investigation. Third, the UMC correction performance is still not perfect, as shown in Figure \ref{fig:cod} blue curves, where the $\Delta$COD values are non-zero. The current implementation uses a simple 3-level UNet for motion prediction. Deploying a more advanced registration network, e.g. transformer-based network \cite{chen2022transmorph} and temporal registration networks \cite{guo2022mcp,guo2022unsupervised,zhou2021mdpet}, may potentially further reduce the registration error and improve the final reconstruction quality. Lastly, the PCI denoising step requires supervised training from paired gated images, which is time-consuming to prepare. In the future, we will also investigate self-supervised denoising methods, e.g. Noise2Void \cite{song2021noise2void}, for PCI denoising in our Fast-MC-PET. 

\vspace{-0.15cm}
\section{Conclusion}
\vspace{-0.25cm}
This paper presents a deep learning-aided motion correction and reconstruction framework for accelerated PET, called Fast-MC-PET. The Fast-MC-PET consisting of UMC and SL-Recon, uses only 2 minutes accelerated PET acquisition data for high-quality reconstruction. The UMC reconstructs motion-corrected short acquisition image, regardless of the motion type in the abdominal region. The SL-Recon then converts the 2 minutes UMC image into virtual 15 minutes UMC image. The experimental results demonstrate that our proposed method can accelerate acquisition by nearly 7-fold and generate high-quality motion-corrected reconstruction for patients with different motions. 

%
%
%
\bibliographystyle{splncs04}
\bibliography{bibliography}

\end{document}